\documentclass[11pt]{article}

\usepackage[dvips]{graphicx, color, xcolor}
\usepackage{epstopdf}

\usepackage[plain]{fullpage}
\usepackage{amsfonts,amssymb,amsbsy,amsmath,amsthm}
\usepackage{subfigure}
\usepackage{verbatim}
\usepackage{fancybox}
\usepackage{multirow} 
\usepackage{cite}
\usepackage{enumerate}

\newcommand{\be}{\begin{equation}}
\newcommand{\ee}{\end{equation}}

\newcommand{\besub}{\begin{subequations}}
\newcommand{\eesub}{\end{subequations}}

\newcommand{\ben}{\begin{equation*}}
\newcommand{\een}{\end{equation*}}

\newcommand\ba{\begin{array}}
\newcommand\ea{\end{array}}
\newcommand{\bea}{\begin{eqnarray}}
\newcommand{\eea}{\end{eqnarray}}

\newcommand{\bean}{\begin{eqnarray*}}
\newcommand{\eean}{\end{eqnarray*}}

\def\@biblabel#1{}

\def\fracD #1{ _0^cD_t^{\alpha({\bf x},t)} {#1} ({\bf x},t)}
\def\fracdd  #1{ _0^cD_t^{\alpha({\bf x},t)} {#1}}

\def\lbd{\lambda}
\def\ka{\kappa}
 \def\ha{\frac 12}
 \def\xt #1{({\bf x},t_{#1})}
 
  \def\bfx #1{{\bf #1}}
 \def \a{\alpha}
 \def\g{\gamma}
 \def\G{\Gamma}
 
  \def \dlt{\Delta}
  
  \def\px #1{\frac{\partial #1}{\partial x}}
    \def\pxx #1{\frac{\partial^2 #1}{\partial x^2}}
        
  \def\py #1{\frac{\partial #1}{\partial y}}
    \def\pyy #1{\frac{\partial^2 #1}{\partial y^2}}

\title
{A novel Hermite RBF-based differential quadrature method for solving two-dimensional variable-order time fractional advection-diffusion equation with Neumann boundary condition
}

\begin{document}
\maketitle
\begin{center}
\author{Jianming Liu$^{a}$ and Xinkai Li$^{b,}$\footnote{Corresponding author. xkl@dmu.ac.uk}\\
\vspace{5mm}
{\small $^a$ School of Mathematics and Statistics, Jiangsu Normal University, Xuzhou 221116, China\\
$^b$ Faculty of Technology, De Montfort University, Leicester LE1 9BH, England
}}
\end{center}

\begin{abstract}
In this paper, a novel Hermite radial basis function-based differential quadrature method (H-RBF-DQ) is presented. This new method is designed to treat derivative boundary conditions accurately. The developed method is very different from the original Hermite RBF method. In order to illustrate the specific process of this method, although the method can be used to study most of partial differential equations, the numerical simulation of two-dimensional variable-order time fractional advection-diffusion equation is chosen as an example. For the general case of irregular geometry, the meshless local form of RBF-DQ was used and the multiquadric type of radial basis functions are selected for the computations. The method is validated by the documented test examples involving variable-order fractional modeling of air pollution. The numerical results demonstrate the robustness and the versatility of the proposed approach.
\end{abstract}
\section {Introduction}
\label{intr_sec}
Currently, numerical methods commonly used in scientific and engineering calculations include finite difference method and finite element method. The finite element method has now become the most widely used numerical method in research and application, and has played a great role in scientific research and engineering analysis.
The finite element method discrete the complex region into elements, and a node based interpolation function approximation is established on the element for the solving variable. The control equation is established by the variational principle or the weighted residual method. It effectively overcomes the limitation of the finite difference method for the shape of the region, and can handle the boundary of various shapes flexibly. However, when the finite element method is used to deal with the problems with characters of large deformation, dynamic crack propagation, fluid solid interaction and so on, the grid can be distorted sometimes. Therefore, mesh reconstruction is unavoidable during the process of numerical simulation, which not only reduces the computational efficiency, but also leads to the damage of the calculation precision. In order to get rid of the dependence on the grid, the meshless method, only based on nodes, has made great progress since 1990s. The structure of the test function of the meshless method is built on a series of discrete nodes. The connection between the field point and the node is no longer realized by the element, so it gets rid of the constraints of the grid or element, and no longer needs the grid reconfiguration when it involves the grid distortion and the grid movement \cite{ChenYM2015Book}. 

To treat the complex computational domains and avoid the mesh generation, meshless techniques have become very popular in recent years.
There are many kinds of meshless methods at present, such as, Smooth particle hydrodynamics method (SPH), Element-free Galerkin method (EFG), Meshless local Petrov-Galerkin method (MLPG), reproducing kernel particle method (RKPM), Radial bases functions method (RBFs), Finite point method (FPM), Moving least square method (MLS), and so on \cite{Liu2005An, Fasshauer2007Book}. Their main difference is the use of different try functions or equivalent forms of differential equations. 
At present, meshless methods can be divided into two categories: weak-form method and collocation method. 
From the weak form variational principle of partial differential equation, the weak form meshless method gives the discrete form of the problem. The characteristic of this method is that the solution is of high accuracy and good stability, but the amount of calculation is high, and the numerical integration is demanded. 
The collocation method directly satisfies the differential equation or boundary condition at discrete points. This method is simple and does not require numerical integration. The computation efficiency is higher than that of the weak form method. 

In 1990, Kansa combined the collocation method with RBFs for computational fluid dynamics \cite{KANSA1990127,KANSA1990147}. After that, there are lots of research on the applications on the RBFs collocation methods in the papers\cite{Shu2003Local,SHU20052001,WRIGHT200699,ROQUE2011363,DEHGHAN2015129,DEHGHAN201723,BAYONA2017257} and their references.  RBFs collocation method is a real meshless method for solving partial differential equations, and it has nothing to do with the spatial dimension \cite{Fasshauer2007Book,DEHGHAN201574}. 
To treat any geometries,  Shu et al. developed a meshfree local RBF-DQ approach to solve the two-dimensional incompressible Navier-Stokes equations \cite{Shu2003Local}.  Radial basis function-based differential quadrature method (RBF-DQ) is a meshless method which has originated from the concept of differential quadrature. Classical differential quadrature (DQ) method began from the idea of conventional integral quadrature \cite{Chang2000Differential}, but it cannot directly be applied to problems with irregular geometries.  Due to the vast flexibility, RBF-DQ method has been successfully used to study many scientific and engineering problems \cite{SHU20052001,HASHEMI20114934, CHAN2014157, DEHGHAN201574, GOLBABAI2016586,GOLBABAI2017130}. 
However, the accuracy achieved by direct meshless collocation scheme is a bit poor especially on boundary \cite{LISZKA1996263, Liu2005Radial}. Furthermore, the collocation scheme has difficulties in dealing with Neumann boundary conditions. In order to improve the accuracy near the derivative boundary, recently, Hermte RBFs methods have been developed by many authors \cite{Liu2005Radial,Liu2005An,Fasshauer2007Book,STEVENS20094606,KROWIAK20162421}. In this paper, we developed a novel Hermite RBF-based differential quadrature method for solving partial differential equations (PDEs). The present method can decrease greatly the cost of computation. The present method can be seen as an extension of meshless local RBF-DQ approach in  \cite{Shu2003Local}, but the accuracy approximating the Neumann boundary conditions can be improved greatly. 
To show the method, we chose a special two-dimensional variable-order time fractional advection-diffusion equation with Neumann boundary condition to demonstrate. 

There are some researches on meshless method for the numerical simulations of time fractional advection-diffusion equation. MLS method is presented to solve the constant-order time fractional advection-diffusion equation in the paper \cite{Zhuang2011,MARDANI2018122}. For the variable-order time fractional advection-diffusion equation, we can refer to the paper \cite{TAYEBI2017655}.  However, the shape function in MLS method is not satisfied with Kronecker $\delta$ function character and  the boundary condition can not be directly enforced, which increase the difficulty for the treatment of boundary. Furthermore, the above works are all on the problem with Dirichlet boundary conditions. 
In recent work, we develop a meshless local RBF-DQ approach to solve the variable-order time fractional advection-diffusion equation. In order to improve the approximating accuracy near the Neumann boundary, in this paper, we develop a novel Hermite RBF-DQ method.

The paper is organized as follows. In Section \ref{timeSec}, the modeling equation and their
time discretization are presented.
The review of RBF-DQ method is presented in Section \ref{fbfdqSec}. For the treatment of the Neumann boundary condition, In Section \ref{bndyTreat}, based on the Hermite RBF interpolation, a novel Hermite RBF-DQ method is described to improve the approximating accuracy.
In Section \ref{sec:numEx}, computational
results  are presented based on two-dimensional variable-order time fractional advection-diffusion equation.

 \section{The modeling equation and time discretization approximation}
 \label{timeSec}
 The variable-order time fractional advection-diffusion equation \cite{TAYEBI2017655} can be formulated as
\be _0^cD_t^{\alpha({\bf x},t)}u({\bf x},t)=\kappa({\bf x},t)\Delta u({\bf x},t)-{\bf v}({\bf x},t)\cdot\nabla u({\bf x},t)+f({\bf x},t),\quad {\bf x}=(x,y) \in\Omega\subset \mathbb{R}^2,\quad t>0, \label{tfade} \ee
subject to the following general initial and boundary conditions
\be\ba{l}
u({\bf x},0)=g({\bf x}),\quad {\bf x} \in \Omega,\\
\mathbb{B}u({\bf x},t)=h({\bf x},t), \quad {\bf x} \in \partial\Omega,\quad t>0,
\ea
\label{initialBoundaryCon}
\ee
where $\Delta u=\pxx u+\pyy u$, $\nabla u=(\px u, \py u)$, and $\mathbb{B}$ denotes the operator with Dirichlet or Neumann boundary conditions.  $\Omega$ is a bounded domain in $\mathbb{R}^2$, $\partial \Omega$ is the boundary of $\Omega$, $\kappa({\bf x},t)>0$ is the diffusion coefficient function, {\bf v}({\bf x},t) is advection velocity, and $f({\bf x},t)$, $g({\bf x})$, $h({\bf x},t)$ are given functions.

In Equation (\ref{tfade}), the variable-order time fractional derivative $\fracD u$ with $0<\alpha({\bf x},t)\leq 1$ in the Caputo definition as 
\be \fracD u=\left\{ \ba{l} \frac{1}{\Gamma(1-\alpha({\bf x},t))}\int_0^t\frac{1}{(t-\xi)^{\alpha({\bf x},t))}} \frac{\partial u({\bf x},\xi)}{\partial \xi} d\xi,\quad 
0<\alpha({\bf x},t)<1, \\
\frac{\partial u({\bf x},t)}{\partial t},\quad\quad\quad\quad\quad\quad\quad\quad\quad\quad\quad\quad
\alpha({\bf x},t)=1.
\ea \right.\label{CapDef}
\ee

 In this study, we focus on the RBF-DQ method for the variable-order time fractional advection-diffusion equation on complex geometries with different boundary conditions. First, in this section, we give the  time discretization approximation.
 Use the time discretization method proposed in the paper \cite{TAYEBI2017655} for Caputo definition, with the notation 
 $b_j({\bf x},t_{k+1})=(j+1)^{1-\a ({\bf x},t_{k+1})}-j^{1-\a ({\bf x},t_{k+1})},$
 we have
 \be\ba{l} \fracdd u({\bf x},t_{k+1})=\frac{1}{\Gamma(1-\alpha({\bf x},t_{k+1}))}
 \sum\limits_{j=0}^{k}\int_{t_{j}}^{t_{j+1}}
 \frac{1}{(t_{k+1}-\xi)^{\alpha({\bf x},t_{k+1})}} 
 \frac{\partial u({\bf x},\xi)}{\partial \xi} d\xi,\\
 \qquad \qquad \qquad  \qquad=\frac{1}{\Gamma(1-\alpha({\bf x},t_{k+1}))}
 \sum\limits_{j=0}^{k}\frac{  u({\bf x},t_{j+1}) - u({\bf x},t_{j}) }{\dlt t} 
 \int_{t_{j}}^{t_{j+1}}\frac{1}{(t_{k+1}-\xi)^{\alpha({\bf x},t_{k+1})}} d\xi +R_{k+1},\\
 \qquad \qquad \qquad  \qquad=\frac{\dlt t^{-\a({\bf x},t_{k+1})}}{\Gamma(2-\alpha({\bf x},t_{k+1}))}\sum\limits_{j=0}^{k}
 b_{k-j}({\bf x}, t_{k+1})(u({\bf x},t_{j+1}) - u({\bf x},t_{j}))+R_{k+1}  \\
 \qquad \qquad \qquad  \qquad=\frac{\dlt t^{-\a({\bf x},t_{k+1})}}{\Gamma(2-\alpha({\bf x},t_{k+1}))}\sum\limits_{j=0}^{k}
 b_{j}({\bf x}, t_{k+1})(u({\bf x},t_{k-j+1}) - u({\bf x},t_{k-j}))+R_{k+1}
     \ea
\label{CapTimeDisc}
\ee
where $t_k=k\dlt t$ for $k=0,1,\dots,M$ and $M=T/\dlt t$. The truncation error $R_{k+1}$ \cite{LIN20071533,TAYEBI2017655,Zhuang2011}
is subjected to 
\be
|R_{k+1}|\leq C \dlt t ^{2-\a ({\bf x}, t_{k+1})}.\label{TruncErr}
\ee
Substituting Equation (\ref{CapTimeDisc}) into Equation (\ref{tfade}), and with the $\theta$-weighted scheme ($\theta\in [0,1]$), we can obtain
\be\ba{l}
u({\bf x},t_{k+1})-\theta\mu\xt{k+1}\left[\ka \xt{k+1} \dlt u\xt{k+1} -{\bf v}\xt{k+1} \cdot \nabla u\xt{k+1}\right] \\
\qquad=u({\bf x},t_{k})+(1-\theta)\mu\xt{k+1}\left[\ka \xt{k+1} \dlt u\xt{k} -{\bf v}\xt{k+1} \cdot \nabla u\xt{k}\right] \\
\qquad - \sum\limits_{j=1}^{k}
 b_{j}({\bf x}, t_{k+1})(u({\bf x},t_{k-j+1}) - u({\bf x},t_{k-j}))+\mu\xt{k+1} f\xt{k+1} +\overline{ R}_{k+1} +\overline{M}_{k+1},
 \ea
\label{thetaWeiSch}
\ee
where 
\be
\mu\xt{k+1}=\dlt t ^{\a\xt{k+1}}\Gamma (2-\alpha({\bf x},t_{k+1})),
\label{muFunc}
\ee
and 
\be
|\overline{ R}_{k+1}  |\leq \overline{C} \dlt t^2,\quad |\overline{ M}_{k+1}  |\leq \overline{C}_1 (1-\theta)\dlt t^{1+\a\xt{k+1}}.
\label{newTruncErr}
\ee

Using the notations $u^k=u^k({\bf x})$ as the numerical approximation to $u\xt{k}$,  $\mu ^k=\mu\xt{k}$, $\ka^k=\ka\xt{k},\  {\bf v}^k={\bf v}\xt{k}=(v_1\xt{k},v_2\xt{k}), \  b_j^k=b_j\xt{k}$  and $f^k=f\xt{k}$, then Equation (\ref{tfade})
can be discretized as follows
\be\ba{l}
u^{k+1}-\theta\mu^{k+1}\left[\ka^{k+1} \dlt u^{k+1} -{\bf v}^{k+1} \cdot \nabla u^{k+1}\right] \\
\qquad=u^k+(1-\theta)\mu^{k+1}\left[\ka ^{k+1} \dlt u^{k} -{\bf v}^{k+1} \cdot \nabla u^{k}\right] 
 - \sum\limits_{j=1}^{k}
 b_{j}^{k+1}(u^{k-j+1} - u^{k-j})+\mu^{k+1} f^{k+1} .
 \ea
\label{timeFinalSch}
\ee
For $\theta=1$, the Equation (\ref{timeFinalSch}) is the same as the scheme proposed by the paper \cite{Zhuang2011} for constant-order  time fractional advection-diffusion equation. The equation (\ref{timeFinalSch}) can be rearranged as
\be\ba{l}
u^{k+1}-\theta\mu^{k+1}\left[\ka^{k+1} \dlt u^{k+1} -{\bf v}^{k+1} \cdot \nabla u^{k+1}\right] \\
\qquad=(1-b_1^{k+1})u^k+(1-\theta)\mu^{k+1}\left[\ka ^{k+1} \dlt u^{k} -{\bf v}^{k+1} \cdot \nabla u^{k}\right] \\
\qquad\quad +\sum\limits_{j=1}^{k-1}
 (b_{j}^{k+1}-b_{j+1}^{k+1})u^{k-j}+b_k^{k+1}u^0+\mu^{k+1} f^{k+1} .
 \ea
\label{timeFinalSch1}
\ee

\section{RBF-based differential quadrature method}
\label{fbfdqSec}
\subsection{Basic RBF-based differential quadrature method}
The idea of differential quadrature method came from the numerical integral, that any integral over a closed domain can be approximated by a linear weighted sum of all the functional values in the integral domain. In differential quadrature method, as the numerical integral, the derivative value $u^{(m)}({\bf x})$ at the centre ${\bf x}_i$ are approximated by a linear weighted sum of the function values at a set of nodes in a closed domain as 
\be
u^{(m)}_x({\bf x_i})\approx \sum\limits_{j=1}^N w_{ij}^{(m)} u({\bf x}_j), \quad
u^{(m)}_y({\bf x_i})\approx \sum\limits_{j=1}^N \overline{w}_{ij}^{(m)} u({\bf x}_j),
\label{grbfdq}
\ee
for $i=1,\cdots,N,$ where  $w_{ij}^{(m)},\   \overline{w}_{ij}^{(m)} $ are the weighting coefficients for derivatives of order $m$ with respect to $x$ and $y$, respectively.
In RBF-DQ method, the weighting coefficients of  $w_{ij}^{(m)},$ and $ \overline{w}_{ij}^{(m)} $ are determined by all the base functions as test function in Equation (\ref{grbfdq}) \cite{Wu2002Development,Shu2003Local}.  

There are many RBFs available. In this study, due to the better performance for the interpolation of 2D scattered data, multiquadrics (MQ) basis function is used as the test function. The function in the region of $\Omega$ can be locally approximated by MQ RBFs as
\be
h(x,y)=\sum\limits_{j=1}^N\lbd_j \sqrt{(x-x_j)^2+(y-y_j)^2+c_j^2} +\lbd_{N+1},
\label{FuncIPbyMQ}
\ee
with shape parameter $c_j$.
To make the problem be well-posed, the equation
\be
\sum\limits_{j=1}^N\lbd_j=0,
\label{lweq}
\ee
is enforced. Substituting Equation (\ref{lweq}) into equation (\ref{FuncIPbyMQ}) gives
\be
h(x,y)=\sum\limits_{j=1,j\neq i}^N\lbd_j g_j(x,y)+\lbd_{N+1},
\label{funcByg}
\ee
where 
\be
g_j(x,y)=\sqrt{(x-x_j)^2+(y-y_j)^2+c_j^2} -\sqrt{(x-x_i)^2+(y-y_i)^2+c_i^2}.
 \label{gbasis}
\ee
The number of unknowns in Equation (\ref{funcByg}) is $N$. As the setting in the paper \cite{Shu2003Local}, $\lbd_{N+1}$ can be replaced by $\lbd_i$, and Equation (\ref{funcByg}) can be written as 
\be
h(x,y)=\sum\limits_{j=1,j\neq i}^N\lbd_j g_j(x,y)+\lbd_{i}.
\label{funcByg2}
\ee
Where $g_i(x,y)=1$ and $g_j(x,y),\  j=1,\cdots,N$ but $j \neq i$ given by Equation (\ref{gbasis}) are a base vector for the function space of $h(x,y)$. 

In RBF-DQ method, the weighting coefficients of  $w_{ij}^{(m)},$ and $ \overline{w}_{ij}^{(m)} $ are determined by all the base functions as the test function in Equation (\ref{grbfdq}), and we can obtain
\besub
\be
\sum\limits_{k=1}^{N}w_{ik}^{(m)}=0,
\ee
\be
\frac{\partial ^m g_j(x_i,y_i)}{\partial x^m}=\sum\limits_{k=1}^{N}w_{ik}^{(m)}g_j(x_k,y_k),\   j=1,2,\cdots,N,\   \mbox{but}\  j\neq i.
\ee
\label{coeBase}
\eesub

The $N$ equations of (\ref{coeBase}) form a linear system with $N$ unknowns for the given $i$, and the weighting coefficients $w_{ik}^{(m)}$ can be solved by a numerical method. In a similar manner, the weighting coefficients $ \overline{w}_{ij}^{(m)} $  of the $y$-derivatives can also be computed by Equation (\ref{coeBase}) with $x$ substituted by $y$.

Substitution of Equation (\ref{grbfdq}) into time discretized equation (\ref{timeFinalSch1}) at point ${\bf x_i}=(x_i,y_i)$,  with $u^{k+1}_i$ as the approximation solution of $u({\bf x}_i, t_{k+1})$, yields
\be\ba{l}
u^{k+1}_i-\theta\mu^{k+1}_i\left[\ka^{k+1}_i \sum\limits_{j=1}^N \left(w_{ij}^{(2)}+\overline{w}_{ij}^{(2)}\right) u_j^{k+1} -\sum\limits_{j=1}^N \left(w_{ij}^{(1)}v_{1,i}^{k+1}+\overline{w}_{ij}^{(1)}v_{2,i}^{k+1}\right)u_j^{k+1}\right] \\
\qquad=(1-b_{i,1}^{k+1})u_i^k+(1-\theta)\mu_i^{k+1}\left[\ka_i^{k+1} \sum\limits_{j=1}^N \left(w_{ij}^{(2)}+\overline{w}_{ij}^{(2)}\right) u_j^{k} -\sum\limits_{j=1}^N \left(w_{ij}^{(1)}v_{1,i}^{k+1}+\overline{w}_{ij}^{(1)}v_{2,i}^{k+1}\right)u_j^{k}\right] \\
\qquad\quad +\sum\limits_{j=1}^{k-1}
 (b_{i,j}^{k+1}-b_{i,j+1}^{k+1})u_i^{k-j}+b_{i,k}^{k+1}u_i^0+\mu_i^{k+1} f_i^{k+1} .
 \ea
\label{fullSch}
\ee
In Equation (\ref{fullSch}), we use the following notations:
\besub
\ben
\mu^{k+1}_i=\mu(\bfx x_i, t_{k+1}), \quad v_{1,i}^{k+1}=v_1(\bfx x_i,t_{k+1}), \quad v_{2,i}^{k+1}=v_2(\bfx x_i,t_{k+1}),
\een
\ben
 \ka^{k+1}_i=\ka(\bfx x_i, t_{k+1}),\quad  b_{i,j}^{k+1}=b_j({\bf x}_i,t_{k+1}), \quad f_i^{k+1}=f(\bfx x_i, t_{k+1}).
\een
\eesub



\subsection{Local RBF-based differential quadrature method}    
When the number of knots, $N$, is large, the coefficient matrix of (\ref{coeBase}) may be ill-conditioned. This limits its application. Hence, we  mainly use the local RBF-DQ method to solve the variable-order time fractional advection-diffusion equation. 

The key of local RBF-based differential quadrature method is that the $m$-order derivatives of $u(\bfx x)$ at $\bfx x_i$ are approximated by the function values at a set of nodes in the neighborhood of $\bfx x_i$ with $N_i$ nodes (including $\bfx x_i$). That is 
\be
u^{(m)}_x({\bf x_i})\approx \sum\limits_{j=1}^{N_i} w_{ij}^{(m)} u({\bf x}_j), \quad
u^{(m)}_y({\bf x_i})\approx \sum\limits_{j=1}^{N_i} \overline{w}_{ij}^{(m)} u({\bf x}_j),\quad i=1,2,\cdots,N.
\label{lrbfdq}
\ee
The corresponding coefficients $w_{ij}^{(m)} $ and $\overline{w}_{ij}^{(m)} $ can be determined by Equation (\ref{coeBase}) with $N_i$ local support nodes in the neighbor of $\bfx x_i$.

Substitution of Equation (\ref{lrbfdq}) into time discretized equation (\ref{timeFinalSch1}) at point ${\bf x_i}$, by the local RBF-based differential quadrature method, yields
\be\ba{l}
u^{k+1}_i-\theta\mu^{k+1}_i\left[\ka^{k+1}_i \sum\limits_{j=1}^{N_i} \left(w_{ij}^{(2)}+\overline{w}_{ij}^{(2)}\right) u_j^{k+1} -\sum\limits_{j=1}^{N_i} \left(w_{ij}^{(1)}v_{1,i}^{k+1}+\overline{w}_{ij}^{(1)}v_{2,i}^{k+1}\right)u_j^{k+1}\right] \\
\qquad=(1-b_{i,1}^{k+1})u_i^k+(1-\theta)\mu_i^{k+1}\left[\ka_i^{k+1} \sum\limits_{j=1}^{N_i} \left(w_{ij}^{(2)}+\overline{w}_{ij}^{(2)}\right) u_j^{k} -\sum\limits_{j=1}^{N_i}\left(w_{ij}^{(1)}v_{1,i}^{k+1}+\overline{w}_{ij}^{(1)}v_{2,i}^{k+1}\right)u_j^{k}\right] \\
\qquad\quad +\sum\limits_{j=1}^{k-1}
 (b_{i,j}^{k+1}-b_{i,j+1}^{k+1})u_i^{k-j}+b_{i,k}^{k+1}u_i^0+\mu_i^{k+1} f_i^{k+1} , 
 \ea
\label{localfullSch}
\ee
for $i=1,2,\cdots, N$.

As shown in the previous subsection, the RBF-DQ approximation of the function contains a shape parameter $c$ that could be knot-dependent and must be determined by the user. In this study, we use the method of normalization of supporting region suggested in the paper of \cite{Shu2003Local} and set $c=5$.

\section{Boundary treatment}
\label{bndyTreat}

The boundary treatment for Dirichlet boundary condition in collocation method is trivial. However, the collocation scheme has difficulties in dealing with Neumann boundary conditions. In this paper, for the variable-order time fractional advection-diffusion equation, we present a new Hermite-type RBF-DQ method for the Neumann boundary condition. 

\subsection{Hermite RBF interpolation}
The approximation of a function  $u({\bf x})$ (see the papers \cite{Liu2005Radial, Liu2005An}) can be written in a linear combination of RBFs at all the $N_i$ nodes (including the $N_i^b$ Neumann boundary points) within the local support domain and the normal derivatives along the normal direction ${\bf n}=(n_x,n_y)$ at the Neumann boundary points as 
\be
u({\bf x})\approx h(x,y)=\sum\limits_{j=1}^{N_i} \lbd_j \phi_j+  \sum\limits_{l=1}^{N_i^b } \g_l \frac{\partial \phi_l^b}{\partial{\bf n}_l}+ \lbd_{N_i+1}, 
\label{nrbfdq}
\ee
where $\lbd_j (j=1,\cdots,N_i),\g_l (l=1,\cdots,N_i^b)$ are interpolation coefficients, and $\phi_j, \phi_l^b$ are radial basis function, respectively. 

In this study, as the Section \ref{fbfdqSec}, the MQ basis function is used as the radial basis function, that is
\be
 \phi_j=\sqrt{(x-x_j)^2+(y-y_j)^2+c_j^2},\quad \phi_l^b=\sqrt{(x-x_l^b)^2+(y-y_l^b)^2+c_l^2},
  \label{mqphi}
\ee
 where ${\bf x} _l^b=(x_l^b,y_l^b)$ is the coordinate for the $l$th normal derivative boundary point.  So we can get
 \be
 \frac{\partial\phi_l^b}{\partial {\bf n}_l}= \frac{\partial\phi_l^b}{\partial x }n_x+ \frac{\partial\phi_l^b}{\partial y}n_y=\frac{(x-x_l^b)n_x+(y-y_l^b)n_y}{\sqrt{(x-x_l^b)^2+(y-y_l^b)^2+c_l^2}}.
  \label{ndmqphi}
\ee

 With the constrains of  (\ref{lweq}) and the notation of (\ref{gbasis}), 
 the equation of (\ref{nrbfdq}) can be reformulated as 
 \be
u({\bf x})\approx h(x,y)=\sum\limits_{j=1,j\neq i}^{N_i} \lbd_j  g_j(x,y)+ \lbd_{i}+  \sum\limits_{l=1}^{N_i^b } \g_l \frac{\partial \phi_l^b}{\partial{\bf n}_l}, 
\label{nrbfdq2}
\ee 
and
\be
\frac{\partial u({\bf x}) }{\partial {\bf n}}\approx \sum\limits_{j=1,j\neq i}^{N_i} \lbd_j  \frac{\partial g_j(x,y)}{\partial {\bf n}}+  \sum\limits_{l=1}^{N_i^b } \g_l \frac{\partial }{\partial {\bf n}}\left(\frac{\partial \phi_l^b}{\partial{\bf n}_l}\right), 
\label{nrbfdq3}
\ee 
 where $g_i(x,y)=1$, $g_j(x,y),\  j=1,\cdots,N_i$ but $j \neq i$ and $\frac{\partial\phi_l^b}{\partial {\bf n}_l},\ l=1,\cdots,N_i^b$ are a base vector for the function space of $h(x,y)$. 
  
Let 
\be
\hat{u}=\left[ u_1 \  \cdots\ u_{N_i}\  \frac{\partial u_1^b}{\partial {\bf n}_1} \  \cdots\  \frac{\partial u_{N_i^b}^b}{\partial {\bf n}_{N_i^b}}\right]^T_{1\times(N_i+N_i^b)},
\label{ue}
\ee
and the coefficients vector 
\be
\hat{a}=\left[ \lbd_1 \  \cdots\ \lbd_{N_i}\  \g_1 \  \cdots\  \g_{N_i^b}\right]^T_{1\times(N_i+N_i^b)},
\label{ue}
\ee
can be obtained by the interpolations at $N_i$ points and the normal derivatives at $N_i^b$  points on Neumann boundary in Equations (\ref{nrbfdq2}) and (\ref{nrbfdq3}). The process can also be expressed by matrix formulation as follows
\be
\hat{u}=\Psi \hat{a}.
\label{uematrix}
\ee
 Thus, the unknown coefficients vector 
 \be
\hat{a}=\Psi^{-1} \hat{u}.
\label{ae}
\ee 
Finally, the approximation form of function can be obtained like
\be
u({\bf x})\approx \phi \hat{a}=\phi\Psi^{-1} \hat{u}=\psi \hat{u},
\label{uappr}
\ee
 where \be\phi=\left[g_1(x,y), \cdots,g_{i-1}(x,y),1,g_{i+1}(x,y), \cdots, g_N(x,y), \frac{\partial \phi_1^b}{\partial{\bf n}_1},\cdots,\frac{\partial \phi_{N_i^b}^b}{\partial{\bf n}_{N_i^b}} \right],
 \label{nbase}
 \ee
 and the matrix of shape functions $\psi$ can be expressed as follows
 \be
 \psi=\left[  \psi_1,\psi_2,\cdots, \psi_{N_i}, \psi_1^H, \psi_2^H,\cdots, \psi_{N_i^b}^H \right].
 \label{psiBase}
 \ee
 Hence, the function $u({\bf x})$ can be approximately expressed as 
 \be
u({\bf x})\approx  \sum\limits_{k=1}^{N_i} \psi_k u_k + \sum\limits_{j=1}^{N_i^b} \psi_j^H \frac{\partial u_{j}^b}{\partial {\bf n}_j}.
\label{uappr2}
\ee

\subsection{Hermite RBF-DQ method}
 From the interpolation formulation (\ref{uappr2}), we can directily get 
  \be
\frac{\partial^m u({\bf x})}{\partial x^m}\approx  \sum\limits_{k=1}^{N_i} \frac{\partial^m \psi_k }{\partial x^m} u_k + \sum\limits_{j=1}^{N_i^b}\frac{ \partial^m\psi_j^H}{\partial x^m} \frac{\partial u_{j}^b}{\partial {\bf n}_j}.
\label{uapprDS}
\ee 
Then the $m$-th order derivative value $u^{(m)}_x({\bf x})=\frac{\partial^m u({\bf x})}{\partial x^m}$ at the centre ${\bf x}_i$ can be approximated by a linear weighted sum of the function values at  $N_i$ points
 and the normal derivatives at $N_b$  points on Neumann boundary  as

\be u^{(m)}_x({\bf x_i})
\approx  \sum\limits_{k=1}^{N_i} w_{ik} ^{(m)}u_k + \sum\limits_{j=1}^{N_i^b} v_{ij}^{(m)} \frac{\partial u_{j}^b}{\partial {\bf n}_j}.
\label{uapprDSpx}
\ee 
In order to get the coefficients in Equations (\ref{uapprDSpx}),  by the idea of differential quadrature in Section \ref{fbfdqSec}, we use all the base functions in (\ref{nbase}) as the test functions and solve the following system
\besub
\be
\sum\limits_{k=1}^{N_i}w_{ik}^{(m)}=0,
\ee
\be
\frac{\partial^m  g_j(x_i,y_i)}{\partial x^m}=\sum\limits_{k=1}^{N_i}w_{ik}^{(m)}g_j(x_k,y_k)+\sum\limits_{l=1}^{N_i^b} v_{il}^{(m)} \frac{\partial g_{j}(x_l^b,y_l^b)}{\partial {\bf n}_l} ,\   j=1,2,\cdots,N_i,\   \mbox{but}\  j\neq i,
\ee
\be
\frac{\partial ^m}{\partial x^m}\left(  \frac{\partial\phi_p^b}{\partial {\bf n}_p}  \right)(x_i,y_i)=\sum\limits_{k=1}^{N_i}w_{ik}^{(m)} \frac{\partial\phi_p^b(x_k,y_k)}{\partial {\bf n}_p} +\sum\limits_{l=1}^{N_i^b} v_{il}^{(m)} \frac{\partial }{\partial {\bf n}_l}\left(\frac{\partial \phi_p^b}{\partial{\bf n}_p}\right)(x_l^b,y_l^b) ,\   p=1,2,\cdots,N_i^b.
\ee
\label{coeBaseDS}
\eesub

In a similar manner, the weighting coefficients $ \overline{w}_{ik}^{(m)},  \overline{v}_{ij}^{(m)} $  of the $m$-th order $y$-derivatives can also be computed by Equation (\ref{coeBaseDS}) with $x$ substituted by $y$. Then the Neumann boundary condition at point $(x_i^b,y_i^b)$ can be formulated as
   \be
\frac{\partial u(x_i^b,y_i^b)}{\partial {\bf n}}\approx  \sum\limits_{k=1}^{N_i} (w_{ik} ^{(1)}n_x+ \overline{w}_{ik}^{(1)}n_y)u_k + \sum\limits_{j=1}^{N_i^b} (v_{ij}^{(1)} n_x+\overline{v}_{ij}^{(1)}n_y)\frac{\partial u_{j}^b}{\partial {\bf n}_j}.
\label{HRBFNB}
\ee 

We must mention that the present Hermite RBF point interpolation method is very different from the ones in \cite{Liu2005Radial, Liu2005An}. The method in  \cite{Liu2005Radial, Liu2005An} firstly required to obtain the matrix of shape functions $\psi$ in (\ref{uappr}); then the differential of components of $\psi$ must be calculated to obtain Equation (\ref{uapprDS}). Furthermore, for time related problems, their method demand that the calculation of the matrix of shape functions $\psi$ with their differential of components should be solved at every time step. In our method, the coefficients in (\ref{uapprDSpx}) is only related to the  
support points set; and for time related problems, we just solve the system of (\ref{coeBaseDS}) once. So the coefficients  in (\ref{uapprDSpx}) can be stored before the time evolution.  

In the present Hermite RBF-DQ method, the normal derivatives at the Neumann boundary nodes are added as additional DOFs and Equations (\ref{HRBFNB}) are enforced. For an internal collocation node at ${\bf x}_i$, if its local support nodes do not include Neumann boundary node, the conventional differential quadrature formulas (\ref{lrbfdq}) are used to approximate the derivatives in Equation (\ref{timeFinalSch1}), and the Equation (\ref{localfullSch}) is used to discretize the  variable-order time fractional advection-diffusion equation.  If its support neighboring nodes include Neumann boundary nodes, the formulation (\ref{uapprDS}) is used.

Although, the derivatives in the Neumann condition can also be discretized by the local RBF-DQ method of (\ref{lrbfdq}) as 
   \be
\frac{\partial u(x_i^b,y_i^b)}{\partial {\bf n}}\approx  \sum\limits_{k=1}^{N_i} (w_{ik} ^{(1)}n_x+ \overline{w}_{ik}^{(1)}n_y)u_k,
\label{LRBFDQNB}
\ee 
the accuracy is far lower than the present Hermite RBF-DQ method, which will be shown in the next section.

\section{Numerical examples}
\label{sec:numEx}
In this section, numerical experiments are carried out to demonstrate the
effectiveness of RBF-DQ method developed in this paper for variable-order time fractional advection-diffusion equation with Dirichlet or Neumann boundary condition. 
Three different test problems are chosen to show the capability and accuracy of the proposed method. These test problems come from the paper \cite{TAYEBI2017655}, but the boundary conditions with Neumann condition and different solution domains are considered. Although Equation (\ref{localfullSch}) is valid for any value of $\theta \in [0,1]$, we use $\theta=1$ as the famous implicit scheme in all numerical examples.

To show the accuracy of the proposed method, the $L^2,\  L^{\infty}$ errors and Root-Mean-Square (RMS) of errors are measured using the following definitions:
\ben
L^2= \sqrt{ \left.{\sum\limits_{i=1}^{N}  \left |    u({\bf x}_i)-u^h({\bf x}_i)  \right |^2} \middle/ { \sum\limits_{i=1}^{N}  \left |  u({\bf x}_i)  \right |^2} \right.},
\een

\ben
L^{\infty}=\max \limits_{1 \leq i  \leq N} \left |  u({\bf x}_i)-u^h({\bf x}_i) \right |, \
 \mathrm{ RMS}= \sqrt{  \frac{1}{N} \left(\sum\limits_{i=1}^{N}  \left |  u({\bf x}_i)-u^h({\bf x}_i)  \right |^2\right)},
\een
where $u^h({\bf x} _i)$ is the numerical solution of  $u({\bf x}_i)$.
\\
\\ {\bf Example 1.} Consider the following variable-order time fractional advection-diffusion equation 
\be
\frac{\partial^{ \alpha({\bf x},t)} u({\bf x},t)}{\partial t^{ \alpha({\bf x},t)}}=\frac{\partial ^2 u({\bf x},t)}{\partial x^2}+ \frac{\partial ^2 u({\bf x},t)}{\partial y^2}-
\frac{\partial u({\bf x},t)}{\partial x}- \frac{\partial  u({\bf x},t)}{\partial y}+
\frac{2t^{2-\a ({\bf x},t)}}{\G(3-\a ({\bf x},t))}+2x+2y-4.
\label{ex1eq}
\ee

\section{Conclusion}
\label{con_sec}


\bibliographystyle{plain}

\bibliography{fadeDQ}{}

\end{document}